\documentstyle[psfig]{mn} 
\title{Photometry of seven overlooked open clusters in the First and Fourth
Galactic Quadrants}
\author[Carraro  at al.]        
{G. Carraro$^{1,2}$,\thanks{Andes Fellow, on leave from Dipartimento
di Astronomia, Universit\`a di Padova,
Vicolo Osservatorio 2, I-35122, Padova, Italy} K. A. Janes$^3$, E. Costa$^1$,
and R. A. M\'endez$^1$,
\thanks{email: 
gcarraro@das.uchile.cl (GC), , janes@bu.edu (KAJ),
costa@das.uchile.cl (EC), 
rmendez@das.uchile cl (RAM)}\\ 
$^1$Departamento de Astronom\'ia, Universidad de Chile, 
Casilla 36-D, Santiago, Chile\\
$^2$Astronomy Department, Yale University, 
P.O. Box 208101, New Haven, CT 06520-8101 , USA\\
$^3$Department of Astronomy, Boston University, 725 Commonwealth Avenue, Boston, MA 02215,
USA\\
 } 
 
\date{\it Submitted: Dec. 2005} 
\pubyear{2005} 
\begin{document} 
\maketitle 
\title{Open clusters in the First and Fourth Galactic Quadrant} 
 
\begin{abstract} 
CCD $BVI$ photometry is presented for 7 previously
unstudied star clusters projected toward the inner
side of the Galaxy: Trumpler~23, Lynga~3,
Collinder~307, Ruprecht~134, ESO552SC16, AL~5 and Kronberger~3.
Color magnitude diagrams of the cluster regions  allow us to conclude
that Lynga 3 and ESO552SC16, are not  clusters, but  groups of bright stars 
probably located in the Carina-Sagittarius 
spiral arm. AL~5 and Kronberger~3 are so embedded in a dense stellar field that we cannot
confirm their nature. Trumpler~23 and Ruprecht~134 are two intermediate-age open clusters
located well inside the solar ring
which deserve further attention.\\
Finally, Collinder~307 is an obscured  younger cluster (250 Myr) located 
in the Carina-Sagittarius spiral arm.
Our results emphasize the difficulty to search for open clusters in the inner
regions of the Galaxy due to the richness of the field and the patchy nature
of the interstellar absorption, but at the same time significantly contribute
to a better understanding of this complicated regions of the Milky Way.
\end{abstract} 
 
\begin{keywords} 
Open clusters and associations: general -- open clusters and associations:  
individual: Trumpler~23, Lynga~3,  Collinder~307, Ruprecht~134 ,ESO552SC16, 
AL~5, Kronberger~3. 
\end{keywords}

\section{Introduction} 
This paper is the third of a series (Carraro et al. 2005a,b)
aimed at increasing the sample of open clusters 
studied in the First and Fourth
Galactic Quadrants. This is a very complicated and barely known
region of the Galaxy due to the highly variable extinction
pattern and the extreme richness of the Galactic disk and bulge fields.\\
In particular the existence and survival of intermediate-age open clusters 
in this region of the Galaxy is very difficult, and indeed
the vast majority of intermediate-age clusters is found in the anti-center
direction (Janes \& Phelps 1994). Inside the solar ring the Galactic tidal
field is very strong and as a consequence star clusters dissolve
more rapidly than in the other regions of the Milky Way disk.
In Carraro et al. (2005a) we discussed NGC~6404 and NGC~6583,
and found that they are intermediate-age clusters located inside the
solar circle. This combination of age and position is very interesting
due to low survival rate of star clusters in these regions.
In Carraro et al. (2005b) we studied 8 candidate clusters, and found that
only 3 are physical
groups, namely AL~1, Berkeley~80 and NGC~5764, thus emphasizing the 
difficulties inherent in studying the stellar populations
toward the inner regions of the galaxy.\\
The final goal of this survey is to find intermediate-age clusters
inside the solar circle,
which will allow us to enlarge the baseline of the radial abundance gradient
holding for the Galactic disk (Friel 2002) and to improve
the open cluster age distribution (Carraro et al. 2005c).
At the same time these clusters, in the event they turn out to be young, 
will help  to better delineate the spiral
structure inside the solar circle.\\
In this paper we present the first CCD photometric study
of seven candidate open clusters, namely Trumpler~23, Lynga~3,
Collinder~307, Ruprecht~134, ESO552SC16, AL~5, and Kronberger~3
with the aim to clarify their nature and provide estimates 
of their fundamental parameters. None of these clusters have
been studied so far, but for Collinder~307 (Moffat \& Vogt 1975).\\
\begin{figure*} 
\centerline{\psfig{file=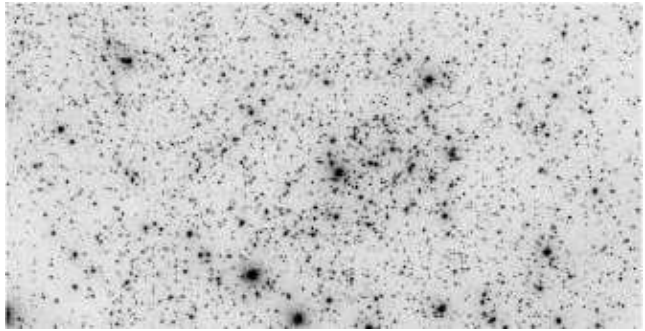}} 
\caption{I band image of Trumpler~23. North is up, East to the left}
\end{figure*}

\begin{figure} 
\centerline{\psfig{file=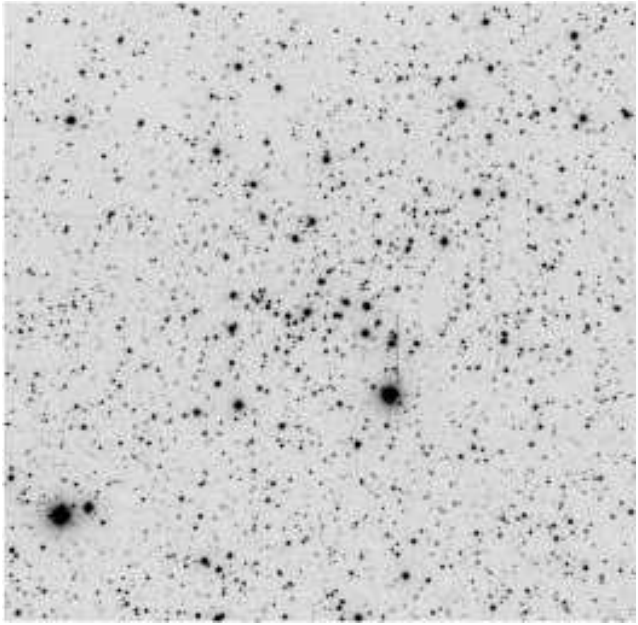,width=\columnwidth,angle=180}} 
\caption{I band image of Lynga~3. North is up, East to the left.}
\end{figure}

\begin{figure} 
\centerline{\psfig{file=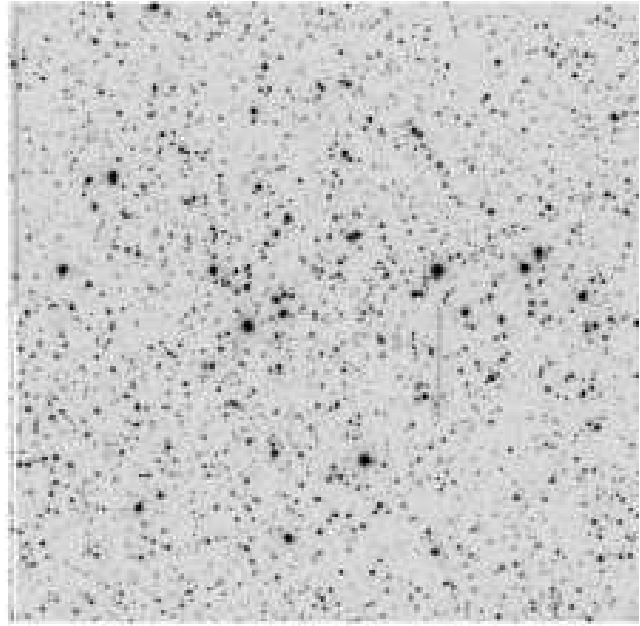,width=\columnwidth,angle=180}} 
\caption{I band image of Collinder~307. North is up, East to the left.}
\end{figure} 

\begin{figure} 
\centerline{\psfig{file=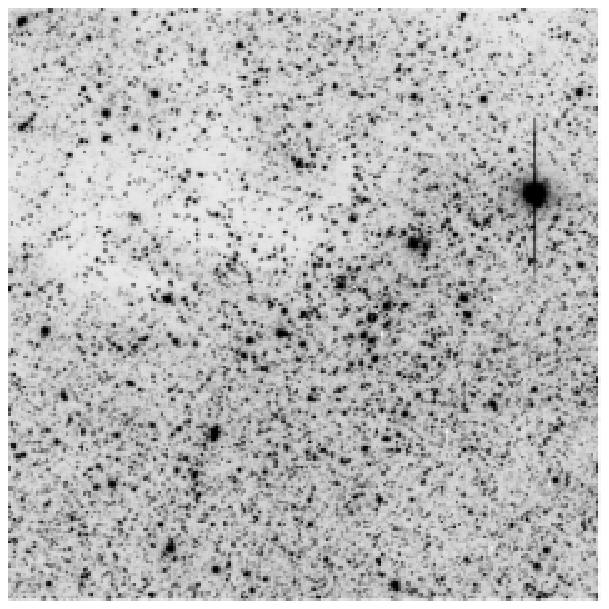,width=10cm,angle=-45}} 
\caption{V band image of Ruprecht~134. North is up, East to the left.}
\end{figure}

\noindent
The layout of the paper is as follows. Sect.~2 illustrates  
the observation and reduction strategies. 
Sect.~3 deals with the Color-Magnitude Diagrams (CMD)
and illustrates the derivation of the clusters' fundamental
parameters. Finally, Sect.~4 provides a detailed discussion
of the results.

 \begin{figure} 
\centerline{\psfig{file=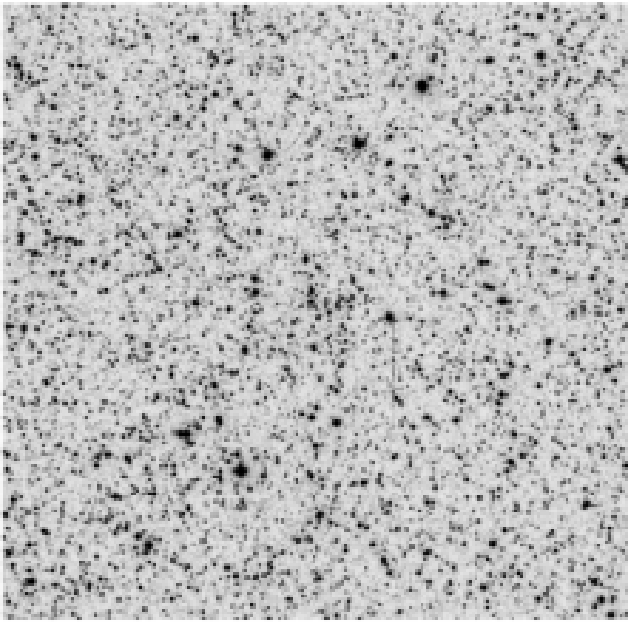,width=\columnwidth,angle=180}} 
\caption{I band image of ESO552SC16. North is down, East to the left.}
\end{figure} 

\begin{figure} 
\centerline{\psfig{file=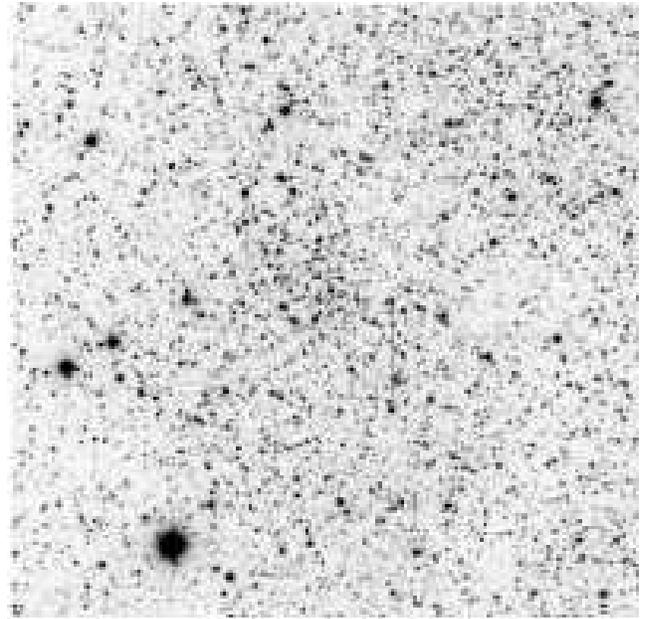,width=\columnwidth,angle=180}} 
\caption{I band image of AL~5. North is up, East to the left.}
\end{figure} 

\begin{figure} 
\centerline{\psfig{file=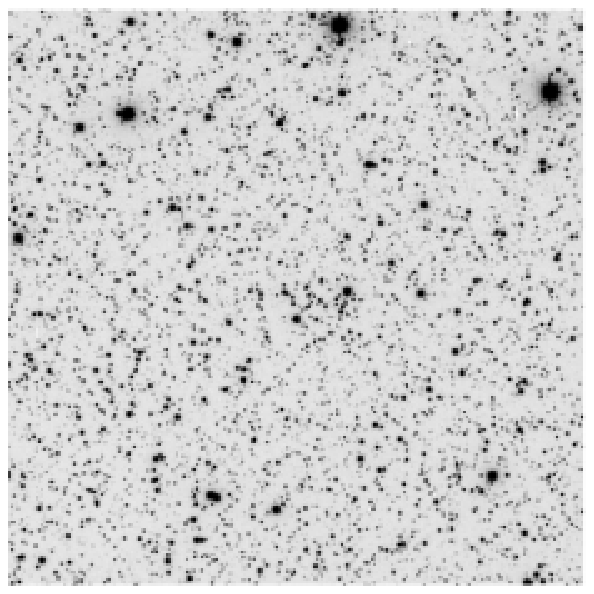,width=10cm,angle=-45}} 
\caption{V band image of Kronberger~3. North is up, East to the left.}
\end{figure} 

\begin{table}
\caption{Basic parameters of the clusters under investigation.
Coordinates are for J2000.0 equinox and have been 
visually re-determined by us.}
\begin{tabular}{ccccc}
\hline
\hline
\multicolumn{1}{c}{Name} &
\multicolumn{1}{c}{$RA$}  &
\multicolumn{1}{c}{$DEC$}  &
\multicolumn{1}{c}{$l$} &
\multicolumn{1}{c}{$b$} \\
\hline
& {\rm $hh:mm:ss$} & {\rm $^{o}$~:~$^{\prime}$~:~$^{\prime\prime}$} & [deg] & [deg]\\
\hline
Lynga~3            & 15:12:36 & -58:08:00 & 321.03 & -0.64\\
Trumpler~23        & 16:00:50 & -53:31:23 & 328.83 & -0.43\\
Collinder~307      & 16:35:20 & -51:00:00 & 334.41 & -2.31\\
Ruprecht~134       & 17:52:43 & -29:33:00 &   0.27 & -1.64\\
ESO552SC16         & 18:19:05 & -27:07:48 &   5.21 & -5.54\\
AL~5               & 18:44:19 & -04:55:48 &  27.73 & -0.67\\ 
Kronberger 3       & 19:39:00 & +06:46:00 &  44.46 & -7.38\\
\hline\hline
\end{tabular}
\end{table}

\section{Observations and Data Reduction} 
The data presented in this paper have been collected during three observing runs 
at Las Campanas (LCO) , Cerro Tololo (CTIO) and La Silla (LSO)
Observatories, Chile. 

\subsection{Las Campanas observations: Trumpler~23}
$\hspace{0.5cm}$
$VI$ observations were carried out with the eight CCD  mosaic camera on-board
the  1. 3m  Warsaw telescope at LCO (Chile), on the night of 
April 4, 2005. 
The cluster was centered in chip $\#$3.
With a pixel size of $0^{\prime\prime}.26$,  and a CCD size of 4096 $\times$ 2048
pixels,  
this samples a $17^\prime.7\times8^\prime.9$ field in the sky.\\

\noindent
Fig.~1 shows the finding chart in the area of Trumpler~23.
North is up, and East on the left.\\
The cluster coordinates were re-determined by us on a visual inspection basis (see Table~1).
\noindent
The data have been reduced with the 
IRAF\footnote{IRAF is distributed by NOAO, operated by AURA under 
cooperative agreement with the NSF.} 
packages CCDRED, DAOPHOT, ALLSTAR and PHOTCAL using the point spread function (PSF)
method (Stetson 1987). 
The night turned out to be photometric, and 
we derived calibration equations for  98 standard stars
observed during the night in the Landolt 
(1992)  fields SA~104-334, PG~0942-029, 
SA110-362, SA~100-269, SA~101-268, G12-43 and   SA~107-601.

\noindent
The calibration equations are of the form:\\

\noindent
$ v = V + (2.206\pm0.005) + (0.15\pm0.02) \times X - (0.033\pm0.006) \times (V-I)$ \\
$ i = I + (2.542\pm0.005) + (0.10\pm0.02) \times X + (0.085\pm0.006) \times (V-I)$ ,\\

\noindent

\noindent
where $VI$ are standard magnitudes, $vi$ are the instrumental ones and  $X$ is 
the airmass. 
The standard 
stars in these fields provide a very good color coverage
being $ -0.5 \leq (V-I) \leq 3.25$.
The final {\it r.m.s.} of the calibration are 0.018 and 0.023 for the V and I filter,
respectively.
Photometric errors have been estimated following Patat \& Carraro (2001, appendix A).
It turns out that stars brighter than  
$V \approx 21$ mag have  
global photometric errors (DAOPHOT internal plus calibration errors)
lower  than 0.20~mag in magnitude and lower than 0.28~mag in colour at V = 21~mag.

\subsection{CTIO observations: Lynga~3, AL~5, Collinder~307 and ESO552SC16}
VI photometry of fields
in the region of Lynga~3, AL~5, Collinder~307 and ESO552SC16
was obtained at the CTIO 1.0m telescope
on June 8  2005, and at the  0.9m telescope
on the nights of 2 and 3 July 2005.
The night of June 8 was not photometric, and we took short (5 sec) and deep
exposures (300 sec). The reduction of these data followed Carraro et al. (2005b) 
and were calibrated using the 0.9m
observations of July 2 and 3.
The pixel scale of the 0.9m 2048 $\times$ 2046 Tek2k $\#$3 CCD
is 0.396$^{\prime\prime}$, yielding a field of 13.5 $\times 13.5$
arcmin on the sky.
The two nights were photometric with an average seeing of 2.0 and 1.5 arcsec,
respectively.
We took only short (10 to 80 secs) exposures
in all the filters to avoid saturation of the brightest stars.\\
Calibration was secured through the observation of Landolt (1992)
standard fields  G~14, G~26, PG~1525, PG~1323, 
SA~110, and SA~112 for a grand total of 86 standard stars.
The calibration equations have the form:\\

\noindent
$ v = V + (2.006\pm0.006) + (0.15\pm0.01) \times X + (0.023\pm0.006) \times (V-I)$ \\
$ i = I + (2.864\pm0.009) + (0.09\pm0.01) \times X + (0.034\pm0.007) \times (V-I)$ ,\\

\noindent
and the final {\it r.m.s} of the calibrated data (fitting plus calibration)
turn out to be 0.022 and 0.025 for the V and I the pass-bands.\\
Stars brighter than  
$V \approx 20$ mag have  
global photometric errors (DAOPHOT internal plus calibration errors)
lower  than 0.30~mag in magnitude and lower than 0.35~mag in colour. \\

\subsection{La Silla observations: Ruprecht~134 and Kronberger~3}
BV photometry of fields
in the region of Ruprecht~134 and Kronberger~3
was taken at the Danish 1.54m telescope
on September 20  2005.  The Danish telescope 2048 $\times$ 2048 CCD
has a pixel scale of 0.395$^{\prime\prime}$, giving 
a field of 13.5 $\times 13.5$
squared arcmin in the sky.
The reduction of this data followed Carraro et al. (2005b).
The night was photometric with an average seeing of 1.5 arcsec.
We took short and deep exposures (10 to 600 secs) exposures
in all the filters to avoid saturation of the brightest stars.\\
Calibration was secured through the observation of Landolt (1992)
standard fields  MarkA, TPhoenix, PG0231 and PG2213 
for a grand total of 56 standard stars.
The calibration equations turned out of be of the form:\\

\noindent
$ b = B + (0.752\pm0.006) + (0.22\pm0.01) \times X - (0.083\pm0.006) \times (B-V)$ \\
$ v = V + (0.280\pm0.006) + (0.18\pm0.01) \times X - (0.005\pm0.006) \times (B-V)$ \\

\noindent
and the final {\it r.m.s} of the calibrated data (fitting plus calibration)
turn out to be 0.032 and 0.025 for the B and V the pass-bands.\\
Stars brighter than  
$V \approx 21$ mag have  
global photometric errors (DAOPHOT internal plus calibration errors)
lower  than 0.30~mag in magnitude and lower than 0.35~mag in colour. \\

\noindent
For all the clusters we performed a completeness analysis,
following the prescriptions described in Carraro et al. (2005d).\\
The final photometric data (X and Y coordinates,
magnitudes and errors)  
consist of 6000, 11225, 7979, 11995, 22729,  13347 and 6553 stars in Lynga~3, Trumpler~23, 
Collinder~307, Ruprecht~134, ESO552SC16, AL~5, and Kronberger~3, respectively, and are
made  available in electronic form at the  
WEBDA\footnote{http://univie.ac.at/webda/navigation.html} site
maintained by E. Paunzen.\\

\begin{figure*}
\centerline{\psfig{file=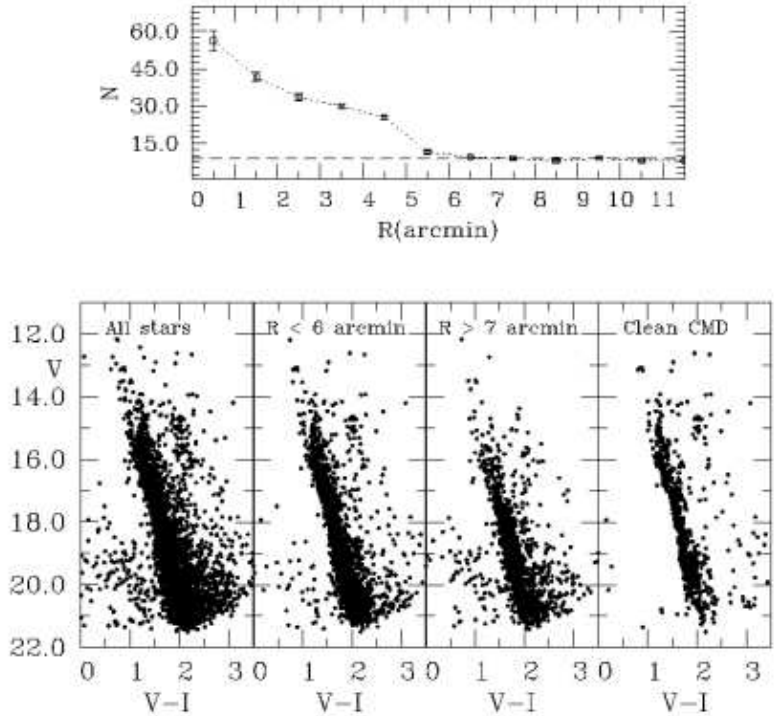}}
\caption{The open cluster Trumpler~23. The upper panel shows
the radial density profile, while the lower panels,
from th left to the right show the CMD of all the observed stars,
and of the cluster, field and and field-decontaminated cluster,
respectively. See text for more details}
\end{figure*}

\section{Colour-Magnitude Diagrams and Cluster Parameters}
In this section we describe cluster CMDs
and derive their basic parameters.

We first evaluated the CMD data as well as images from the 2-Micron All Sky
Survey (2MASS) all sky data release (available at
www.ipac.caltech.edu/2mass/releases/allsky) to explore the
existence of the clusters as physical systems.  The 2MASS $K_s$\ images are
substantially less affected by reddening than the visual images which means
that in some cases, the confusion from background galactic stars can be higher,
but the background should often also be less variable.

Distance moduli, reddenings  and ages of the confirmed clusters have been
derived by matching by eye the
observed CMDs to isochrones from the Padova group (Girardi et al. 2000), 
paying
particular attention to the shape of the Main Sequence (MS), the position of the brightest
MS stars, the turn-off point (TO) and the mean location of evolved stars, if
present.\\
To infer the heliocentric distances we adopted $R_V=A_V/E(B-V) = 3.1$.\\

The results are summarized in Table~2, where the basic parameters
are listed together with their uncertainties. The latter correspond to the
shift allowed to isochrone fitting before a mismatch is clearly perceived
by eye inspection.\\

\begin{figure}
\centerline{\psfig{file=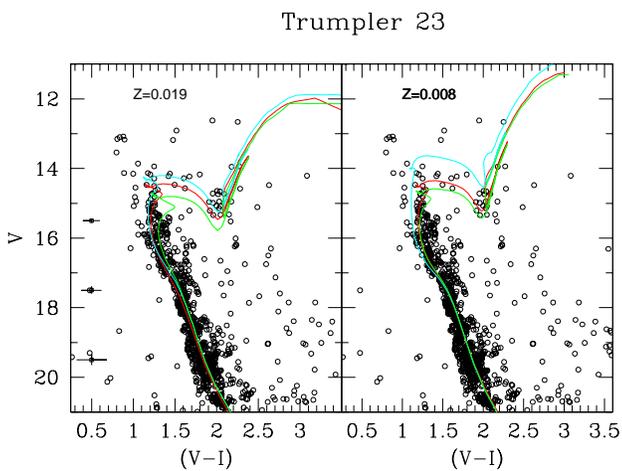,width=\columnwidth}}
\caption{Isochrone solution for Trumpler 23.
In the left panel isochrones of ages of 0.8, 0.6 and 0.5
Gyr (from the bottom to the top) and solar metallicity are shown.
In the right panel isochrones of ages of 0.8, 0.7 and 0.5
Gyr (from the botton to the top) and for Z=0.008 are shown.
See text for more details.}
\end{figure}

\subsection{Trumpler~23}
Trumpler ~23 
was firstly noted by Trumpler (1929) and then by van den Bergh \& Hagen (1975),
who named it BH~180, and
described it as a moderately populated cluster having
a diameter of 6 arcmin, clearly visible both on red and on blue
plates (see Fig.~1). 
To assess the reality of this clear over-density, we statistically
cleaned the CMD following the prescriptions described in
Bertelli et al. (2003).
The procedure is shown in Fig.~8. The left panel presents the CMD
of all the stars observed in the region of Trumpler 23.
Here the
MS extends from $V$=15.5 to $V$=22.0,
with a TO located at $V\approx$16.0, $(V-I)\approx$1.1.\\
This MS is rather broad, substantially wider
than the photometric errors at a given magnitude (see Sect.~2).
We ascribe this to
field star contamination, to some differential reddening
and perhaps to the presence of a sizeable binary star population,
which mainly enlarges the MS toward red colors.
A significant over-density of stars at V $\approx$ 15, (V-I)
$\approx$ 2 might be the signature of a  Red Giant Branch (RGB)
clump of He-burning stars.\\
However the assessment of the reality of this feature is
complicated by field star contamination.\\

\noindent
The first step of the cleaning procedure is to select probable cluster members
according to the distance from the nominal cluster center.
To this aim we performed a star count analysis, counting the number
of stars falling in concentric area-normalized anuli 0.5 arcmin wide, centered
on the cluster center. The result is shown in the upper panel
of the same figure. Here the dashed line is the level of the field
estimated from star counts in a region 7 arcmin apart
from the cluster center. Interestingly we find that the cluster is larger
than previous estimates, and suggest that the radius of Trumpler 23
is about 5.0 arcmin, and it is defined as the distance from the cluster
center where star counts reach the field level.
The shape of the radial density profile is already an evidence
of the cluster reality.
The second step is to consider as probable cluster members
all the stars falling inside the cluster radius (mid-left panel), and as field
stars all the stars lying beyond 7 arcmin from the cluster center
in a equal area field (mid-right panel).
Now, we proceed to clean the CMD, by adopting the following technique.
For any field star in the field CMD, we look for the closest cluster
star in the cluster CMD, and remove it from the cluster CMD.
This procedure in principle takes into account the different
field and cluster completeness, although in this case this is not
an issue, being cluster and field inside the same frame.
The result is shown in the right panel of Fig.~8.

The following considerations can be made:

\begin{itemize}
\item The MS and the TO region in this panel are
much better defined; in particular the MS is quite narrow indicating a modest
field star contamination down to $V \approx 20.0$.
A few stars located above the TO could be Blue Stragglers,
common in open star clusters (Ahumada \& Lapasset 1995), or interlopers;
\item  There is  a nice clump of stars at
$V$=14.80, $(V-I)$=2.0, similar to the clump observed in  open clusters
like NGC~2477 (Kassis et al. 1997) or NGC~6583 (Carraro et al. 2005a).
\item The fine structure of the MS deserves further attention,
since a binary stars may concur to blur this region.
\end{itemize}
Finally, the detailed shape of the TO deserves some attention. In fact
the shape of the TO is that one typical of intermediate-age open clusters,
with red hook (the MS termination point) clearly visible, 
notwithstanding some field
and binary star
contamination.
Again, the shape of the TO and the presence of a clump indicate an age
in the range 0.5-1.5 Gyr, depending on the metallicity.
A preliminary estimate of the cluster age can be derived by using the 
Carraro \& Chiosi (1994)
calibration, which is based on the magnitude difference $\Delta V$ between
the TO and the mean clump magnitude. In this case
$\Delta V$ is about 0.8-1.0 mag., which implies an age around 1 Gyr,
depending on the metallicity of Trumpler~23.\\

\noindent
To derive the cluster fundamental parameters, we make use of the comparison
between the stars distribution in the CMD and a set of theoretical isochrones
from the Padova group (Girardi et al. 2000).
We already have an indication of the cluster age, but we lack information
about the reddening, distance  and the metallicity.

The results of the fits are shown in Fig.~9.
\noindent
In the left panel
we present Trumpler~23 clean CMD 
and super-impose isochrones of 0.5, 0.6 and 0.8 billion
years for a solar (Z=0.019) metallicity. 
The fit is quite good both along the MS,
in the TO and in the evolved stars region for the 0.6 billion years
isochrone. \\
We achieved this results by shifting the isochrone with E$_{(V-I)}=1.05\pm0.05$
(E$_{(B-V)}=0.84$), and $(m-M)_V=14.35\pm0.20$ (errors by eye).\\
The right panel shows our fit for Z=0.008, and for ages
of 0.5, 0.7 and 0.8 Gyrs. The best fit is achieved with the 0.7 billion
years isochrone, and implies
 E$_{(V-I)}=1.15\pm0.05$
(E$_{(B-V)}=0.92$), and $(m-M)_V=14.25\pm0.20$.
The fits in the two panels are comparably good, and therefore
we conclude that the cluster age is about $0.7\pm0.1$ Gyrs,
while the cluster reddening is 1.10$\pm$0.10.
As for the distance modulus, we adopt 14.30$\pm$0.20.\\
\noindent
As a consequence, Trumpler~23 turns out to be located about 2.2 kpc from 
the Sun, in the Fourth Galactic Quadrant.
This implies a distance from the Galactic
center of 6.7 kpc and a height of 16 pc below the Galactic plane. \\

\subsection{Lynga~3}
This cluster  was first noted by Lynga (1964).
It is identified by a concentration of a few bright stars (see Fig.~2) in the direction
of the Carina spiral arm. 
We inspected the 2MASS catalogue to see whether in the $K_s$
image an overdensity appears, but this 
does not show any cluster at the limiting magnitude of the image.
The CMD of the apparent over-density
(Fig.~10, left panel) does not show any significant difference 
with respect  to a comparison field (Fig.~10, mid panel). The putative
MS has the same mean color and slope of the field apart
from a vertical group of stars at (V-I) $\approx$ 2 
and $13.8 \leq V \leq 16.0$. This is readily visible also in the right
panel of the same figure, where the dotted line is the cluster 
histogram, and the solid line the field histogram. The two
histogram are indeed identical but for the aforementioned
group of stars.\\
We argue that Lynga~3 is not a cluster, but an
over-density of a few
bright stars probably  located inside the Carina arm. 

\begin{figure}
\centerline{\psfig{file=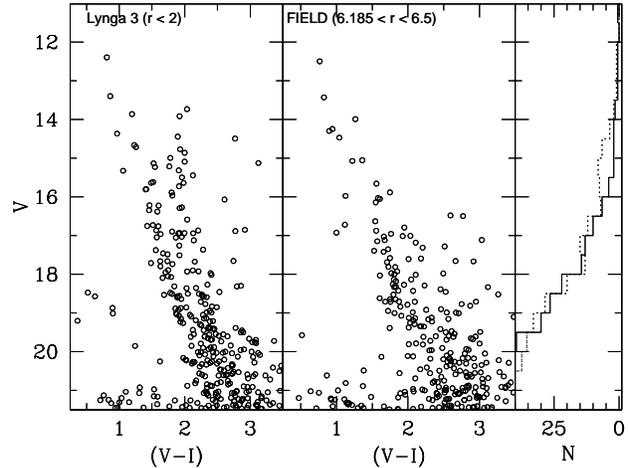,width=\columnwidth}}
\caption{$V$ {\it vs} $(V-I)$ CMDs
of Lynga~3 as a function of  radius from the adopted cluster center.
{\bf Left panel}: the cluster area, derive inspecting Fig.~2.
{\bf Middle panel}: the field area.{\bf Right panel}:
histograms of the cluster (dotted line) and the field (solid line)}
\end{figure}

\begin{figure}
\centerline{\psfig{file=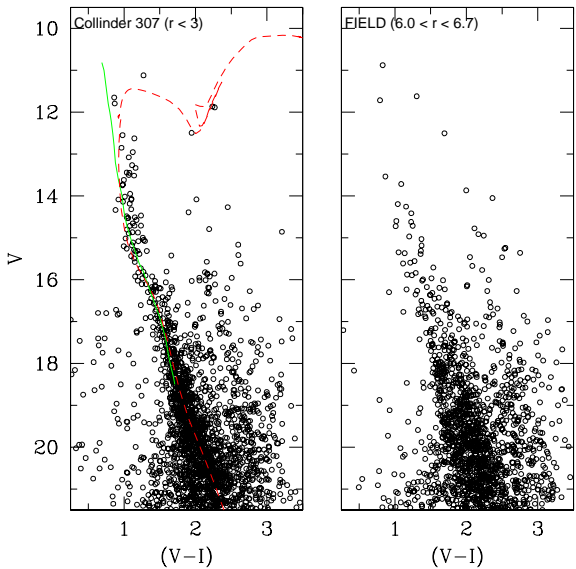,width=\columnwidth}}
\caption{$V$ {\it vs} $(V-I)$ CMDs
of Collinder~307 as a function of  radius from the adopted cluster center.}
\end{figure}

\begin{figure*}
\centerline{\psfig{file=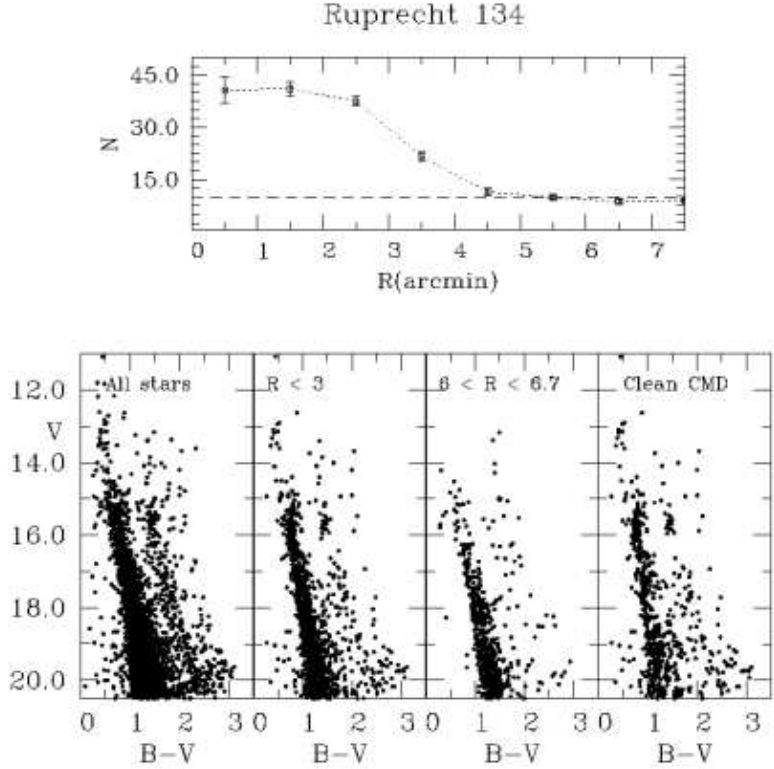}}
\caption{The open cluster Ruprecht~134. The upper panel shows
the radial density profile, while the lower panels,
from th left to the right show the CMD of all the observed stars,
and of the cluster, field and and field-decontaminated cluster,
respectively. See text for more details.}
\end{figure*}

\subsection{Collinder~307}
This cluster was discovered by Collinder (1931)
and then by van den Bergh \& Hagen (1975),
who named it BH~193, and
described it as a poorly populated cluster having
a diameter of 6 arcmin, clearly visible both on red and on blue
plates (see Fig.~3). Moffat \& Vogt (1975) obtained UBV photometry of
a few stars, which turned out to be red, and concluded that there
was no evidence of a star cluster in that direction.
The 2MASS $K_s$
image we inspected 
shows beyond any doubt a sparse cluster at the limiting magnitude of the image.
We construct the CMD as a function of the distance
from the adopted cluster center, although it is clear
from Fig.~3 that the cluster occupies all the region we observed,
and for this reason we are not going to provide a radial density
profile. 
Left panel of Fig.~11
shows the cluster area, namely all the stars lying within
3 armin from the cluster center, whereas the right panel shows a blank field
extracted from the outskirts of the CCD image.
The cluster clearly exists, with a distinct MS extending
from V = 13 down to V = 20. The slope of the MS
demonstrates that the cluster is relatively young.
We superposed an empirical Schmidt-Kaler(1982) ZAMS (solid line)
which fits the bulk of the stars adopting E(V-I) = 1.05 (E(B-V) = 0.84) 
and $(m-M)_V$ = 13.4 .
This implies that the cluster lies at 1.3 kpc from the Sun, and 60
pc below the Galactic plane, well inside the Carina-Sagittarius
spiral arm.
A solar metallicity isochrone shifted by the same reddening and distance
modulus (dashed lines) yields a nice fit for an age of 250 Myr.

\subsection{Ruprecht~134}
This clustering is reported by Ruprecht (1966) who assigns to it
Trumpler (1929) type {\it III 1 p}, say a poorly
populated sparse open cluster (see Fig.~4). In this figure the cluster
appears as a clump of bright stars. The northern part of the field
is dominated  by a conspicuous dust lane.
We inspected also for this cluster a 2MASS $K_s$ image,
and actually we can see a sparse distribution of a few bright stars,
which indicates a possible star cluster.\\

\noindent
To assess the cluster reality we employ the same technique
we used for Trumpler~23, and the results are presented in Fig.~12.\\
The left panel presents the CMD of all the stars
in the covered field.
We built up the cluster CMD considering the stars located
within 3 arcmin from the cluster nominal center (Table~1)
and estimating the field star contamination in a equal
area ring 
located in the outskirts of the surveyed field. This choice
was driven by the appearance of the cluster radial density profile
(see the upper panel of Fig.~12): it is readily seen that the cluster
reaches the level of the field at a distance of about 4.0-4.5 arcmin.

The two CMDs
are shown in the middle-left  and middle-right panels of Fig.~12.
The CMD of the field decontaminated cluster is shown in the right
panel of the same figure.
This panel shows a tight MS and a clump 
of stars at V =15.5, (B-V) = 1.5. This clump in the middle-right panel
is not as visible as the cluster clump, and the MS is
more scattered and terminates
1.5 mag. fainter than in the cluster. We interpret these facts
as signatures of the existence of a intermediate age star
cluster, very similar to Trumpler~23 (see Fig.~8).\\
\noindent
To derive the cluster fundamental parameters
we super-pose a  solar metallicity
isochrone, shifted by E(B-V) = 0.5 and $(m-M)_V$ = 14.2, which provides
a reasonable fit of the TO and clump region for an age
of 1.0 Gyr (see Fig.~13). 
We tried younger ages to better match the clump position.
However by decreasing the age only by a few million years
the theroretical clump  gets too red compared with the observed
one, and for this reason we exclude ages significantly
younger than 1.0 Gyrs.
In turn, this yields a distance of 3.4 kpc
from the Sun exactly toward the Galactic Center, and a
Galactocentric distance of 5.2 kpc.
As for Trumpler~23, we are facing here a really interesting object,
which deserves further attention due to its unusual combination
of age and position in the Galaxy.

\begin{figure}
\centerline{\psfig{file=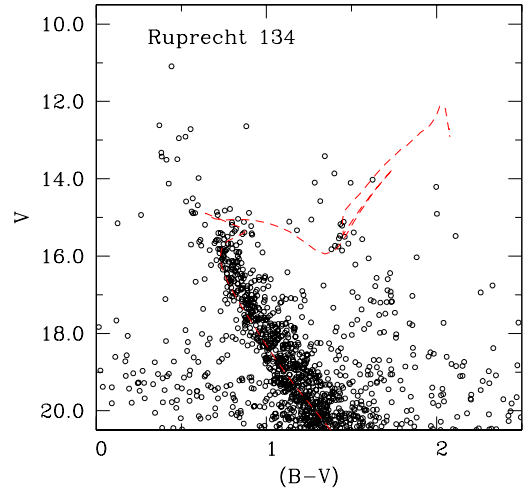,width=\columnwidth}}
\caption{Isochrone solution for the cluster
Ruprecht~134. The 1 Gyr solar metallicity isochrone has been
shifted by E(B-V)=0.5 and $(m-M)_V$=14.2}
\end{figure}

\begin{figure}
\centerline{\psfig{file=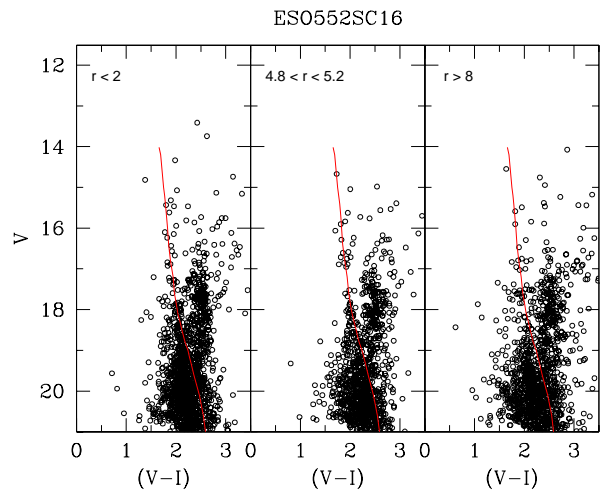,width=\columnwidth}}
\caption{$V$ {\it vs} $(V-I)$ CMDs
of ESO552SC16 as a function of  radius from the adopted cluster center.
The superposed ZAMS has been shifted by $(m-M)_V$ = 16.6 and E(V-I)=1.90}
\end{figure}

\subsection{ESO552SC16}
This asterism is  included in the ESO/Uppsala catalogue (Lauberts 1982)
and is a very compact clustering embedded in a rich stellar field (see Fig.~5).
The 2MASS $K_s$
image which we inspected
does not show any cluster at the limiting magnitude of the image.
The CMD of the apparent over-density
(Fig.~14, left panel) does not show any significant difference 
when selecting different areas of the CCD frame.
We conclude that ESO552SC16 is not a physical cluster.\\
\noindent 
Interestingly, we detect an almost vertical sequence at (V-I)$\approx$
2.0, 19 $\leq V \leq$ 15. This is clearly a sequence of young stars
evenly distributed across the covered field.
A fit with the empirical Schmidt-Kaler ZAMS yields a heliocentric distance 
of 2.2 kpc ( E(V-I) = 1.9 and $(m-M)_V$ = 16.6) and a height below the 
plane of about 200 pc. The Galactic position of this young population is compatible
with the pattern of the Carina-Sagittarius spiral arm.
The apparent over-density in the direction of  ESO552SC16 which
looks like a compact cluster is probably
an extinction effect.

\subsection{AL~5}
This cluster was discovered by Andrews \& Lindsay (1967) 
who describe it as a circular ensemble of faint stars
with a diameter of 2.25 arcmin (see Fig.~6). 
The 2MASS $K_s$ image does not show
a cluster at the limiting magnitude, and the DSS B image neither.
The impression of a cluster is the results of the patchy extinction
pattern in the region.
We then build up CMDs as a function of stars' distance from the 
nominal cluster center. 
The CMD (see Fig.~15) of the apparent over-density
(left panel) does not show any significant difference 
with respect  to a comparison field (right panel). The putative
Main Sequence has the same mean color and slope of the field.
We believe that the small difference in the number of stars
is simply an absorption effect. We conclude that there is no cluster
in the direction of AL~5.

\begin{figure}
\centerline{\psfig{file=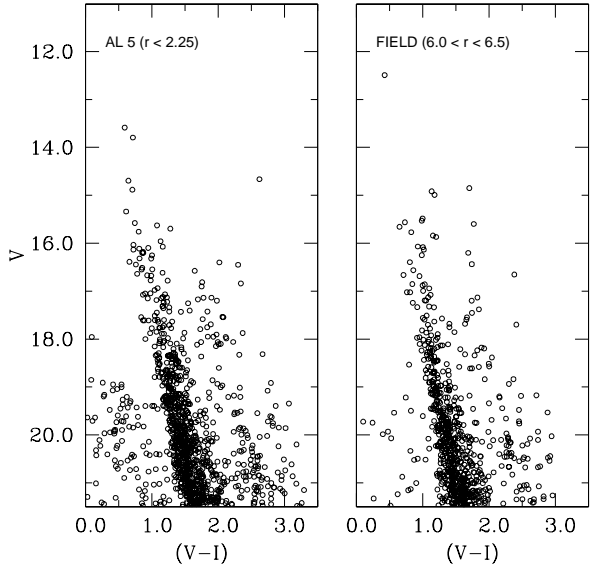,width=\columnwidth}}
\caption{$V$ {\it vs} $(V-I)$ CMDs
of AL~5 as a function of  radius from the adopted cluster center.
The CMDs refer to two equal area fields.}
\end{figure}

\begin{table*}
\caption{Parameters of the studied clusters. The coordinate system
is such that
the Y axis connects the Sun to the Galactic Center, while the X axis is
positive in the direction of galactic rotation.
Y is positive toward the Galactic anti-center,
and X is positive in the first and
second Galactic quadrants (Lynga 1982).}
\fontsize{8} {10pt}\selectfont
\begin{tabular}{ccccccccc}
\hline
\multicolumn{1}{c} {$Name$} &
\multicolumn{1}{c} {$E(B-V)$}  &
\multicolumn{1}{c} {$(m-M)_V$} &
\multicolumn{1}{c} {$d_{\odot}$} &
\multicolumn{1}{c} {$X_{\odot}$} &
\multicolumn{1}{c} {$Y_{\odot}$} &
\multicolumn{1}{c} {$Z_{\odot}$} &
\multicolumn{1}{c} {$R_{GC}$} &
\multicolumn{1}{c} {$Age$} \\
\hline
& mag & mag& kpc & kpc & kpc & pc & kpc & Myr \\
\hline
Trumpler~23    & 0.85$\pm$0.05  & 14.3$\pm$0.2 & 2.2 &  -1.1&   -1.9 &  -16 & 6.7&  1000$\pm$200\\
Lynga~3        & Spiral arm  &  & &  &   &  &  &  \\
Collinder~307  & 0.84$\pm$0.10  & 13.4$\pm$0.2 & 1.3 &  -1.6&   -0.8 &  -60 & 6.9&  250$\pm$50\\
Ruprecht~134   & 0.50$\pm$0.05  & 14.2$\pm$0.2 & 3.4 &   0.0&   -3.4 & -100 & 5.2&  1000$\pm$200  \\
ESO552SC16     & spiral arm  &   &  &    &    &    &   &  \\
AL~5           & No cluster  &   &  &    &    &    &   &  \\
Kronberger~3   & No cluster  &   &  &    &    &    &   &  \\
\hline
\end{tabular}
\end{table*}

\subsection{Kronberger~3}
This small clustering  (see Fig.~7) has been identified by Matthias Kronberger,
who already suggested to be an asterim, and not a real cluster (private communication
to Wilton Dias,\footnote{http://www.astro.iag.usb.br/~wilton/removed.txt}). Both the inspection of the 2MASS image and the global
appearance of the CMD (see Fig.~16) strengthen the impression that Kronberger~3
is just a random group of stars. In fact the CMD of the putative cluster
does not differ from the that of surrounding field.
We therefore conclude that Kronberger~3 is not an open cluster.

\begin{figure}
\centerline{\psfig{file=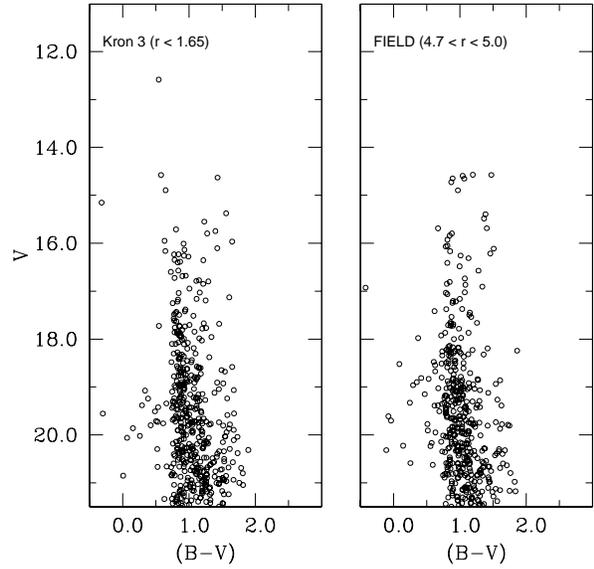,width=\columnwidth}}
\caption{$V$ {\it vs} $(V-I)$ CMDs
of Kronberger~3 as a function of  radius from the adopted cluster center.
The CMDs refer to two equal area fields.}
\end{figure}

\section{Discussion and Conclusions}
The derived parameters of the program clusters are listed in
Table~2.
Together with reddening, distance, age and the corresponding uncertainties,
we list the Galactocentric distance, derived by assuming $R_{\odot}=8.5 kpc$
and the Galactocentric rectangular coordinates $X_{\odot}$, $Y_{\odot}$ and
$Z_{\odot}$.
The adopted reference system is centered on the Sun,
with the X and Y axes lying on the Galactic plane
and Z perpendicular to the plane.
X points in the direction of the Galactic rotation, being positive in the first
and second Galactic quadrants; Y points toward the Galactic anti-center,
being positive in
the second and third quadrant; finally,
Z is positive toward the north Galactic pole (Lynga 1982).\\

Trumpler~23 and Ruprecht~134 are two very interesting clusters of intermediate age located inside the solar ring
which, together with NGC~6404 and NGC 6583 (Carraro et al 2005a) and AL~1 (Carraro et al. 2005b),
significantly increase the number of open clusters of that age known to lie so close 
to the Galactic center. Future spectroscopic programs should consider these clusters
since they would allow to enlarge the baseline of the Galactic disk radial abundance gradient by more
than 2 kpc.\\
Furthermore, Trumpler~23 (see Fig~1) has an elongated shape which might indicate that it is
undergoing strong tidal interaction with Milky Way.\\

Collinder~307 results to be a 250 Myr old cluster located 100 pc below the Galactic plane,
in the Carina-Sagittarius spiral arm.

All the other clusters are just visual effects, accumulations of stars produced
by the patchy nature of the interstellar absorption toward the Galactic bulge.

This work highlights the difficulties of working with open clusters toward
the inner regions of the galaxy.  The star densities are large, and increase
rapidly with distance.  That causes the appearance of a main sequence on all
of the CMDs in this paper, resulting simply from the geometry of the situation.
Furthermore, patchy obscuration is likely to play a role in creating
apparent ``clusters'' that are not physical associated groups of stars.

Nevertheless, in recent works (Carraro et al 2005a,e) we have discovered
a considerable number of neglected intermediate-age open clusters,
which are going to significantly modify the open cluster age distribution
and probably the typical open cluster lifetime as presently known.
 
\section*{Acknowledgements} 
The observations presented in this paper have been carried out at 
LCO, CTIO and LSO, Chile. 
CTIO is operated by the Association of Universities for Research in Astronomy,
Inc. (AURA), under a cooperative agreement with the National Science Foundation
as part of the National Optical Astronomy Observatory (NOAO).
The work of G. Carraro is supported by {\it Fundaci\'on Andes}.
R. A. M\'endez and E. Costa acknowledge support from the
Chilean {\sl Centro de Astrof\'\i sica} FONDAP No. 15010003.
G. Carraro thanks Ricardo Salinas and Marcio Catelan for securing part of the observations
described in this paper
This study made use of Simbad and WEBDA databases.

\end{document}